\documentclass[aps,twocolumn,floatfix,superscriptaddress,aps,longbibliography]{revtex4-1}
\usepackage{epsfig}
\usepackage{amsfonts}
\usepackage{latexsym}
\usepackage{amsmath}
\usepackage{amssymb}
\usepackage{slashed}
\usepackage{longtable}
\usepackage{mathrsfs} 
\usepackage{float}
\usepackage[normalem]{ulem}

\usepackage{graphicx}

\newcommand{\bea}{\begin{eqnarray}}
\newcommand{\eea}{\end{eqnarray}}
\newcommand{\be}{\begin{equation}}
\newcommand{\ee}{\end{equation}}
\newcommand{\ba}{\begin{align}}
\newcommand{\ea}{\end{align}}

\usepackage[colorlinks=true,citecolor=blue,linkcolor=magenta]{hyperref}

\usepackage{xcolor}

\newcommand{\ak}[1]{{\color{black}{{#1}}}}
\newcommand{\rub}[1]{{\color{black}{{#1}}}}

\begin{document} 

\title{Bose-Hubbard realization of fracton defects}

\author{Krzysztof Giergiel}
\affiliation{Institute of Theoretical Physics, Jagiellonian University, 30-348 Krak\'{o}w, Poland}
\author{Ruben Lier} 
\affiliation{Max Planck Institute for the Physics of Complex Systems, 01187 Dresden, Germany}
 \author{Piotr Sur\'owka} 
\email{piotr.surowka@pwr.edu.pl}
\affiliation{Max Planck Institute for the Physics of Complex Systems, 01187 Dresden, Germany}
\affiliation{Department of Theoretical Physics, Wroc\l{}aw  University  of  Science  and  Technology,  50-370  Wroc\l{}aw,  Poland}
\affiliation{Institute for Theoretical Physics, University of Amsterdam, 1090 GL Amsterdam, The Netherlands}
\affiliation{Dutch Institute for Emergent Phenomena (DIEP), University of Amsterdam, 1090 GL Amsterdam, The Netherlands}
\author{Arkadiusz Kosior} 
\email{arkadiusz.kosior@uibk.ac.at}
\affiliation{Institute for Theoretical Physics, University of Innsbruck, 6020 Innsbruck, Austria}

\date{\today}

\begin{abstract}
Bose-Hubbard models are simple paradigmatic lattice models used to study dynamics and phases of quantum bosonic matter. We combine the extended Bose-Hubbard model in the hard-core regime with ring-exchange hoppings. By investigating the symmetries and low-energy properties of the Hamiltonian we argue that the model hosts fractonic defect excitations. We back up our claims with exact numerical simulations of defect dynamics exhibiting mobility constraints.  Moreover, we confirm the robustness of our results against 
fracton symmetry breaking perturbations. Finally we argue that this model can be experimentally realized in recently proposed quantum simulator platforms with big time crystals, thus paving a way for the controlled study of many-body dynamics with mobility constraints. 
     \end{abstract}
\date{\today}

\maketitle

\section{Introduction}
Interacting many-body systems exhibit a variety of collective phenomena and different phases of matter. They are usually accompanied by emergent quasiparticles, whose properties differ from the elementary excitations. One example of such a behavior is given by fractons - emergent quasiparticles that lack the ability to move, either completely or partially. Models with such excitations fall into two distinct categories: gapless and gapped.  Gapless excitations appear in higher-rank gauge theories that emerge in the description of spin-liquids \cite{Xu2006,Xu2010,Pretko2017spinliquid,Pretko2017spinliquid2,You_emergent_2020},  dipole-conserving lattice models \cite{Pai2019,GromovLucas2020,Feldmeier2020,Morningstar2020,Iaconis2021,moudgalya2021spectral}, elasticity \cite{PretkoSolid2018,Gromov2019elastic,Kumar2019,Pretko:2018fed,pretko2019crystal,zhai2019two,Gromov:2019waa,Nguyen:2020yve,Manoj:2020abe,Surowka:2021ved} and hydrodynamics \cite{Doshi2021,Grosvenor:2021rrt,Glorioso:2021bif}. In the gapless theories fractons can be understood as charges that act as sources to the gauge fields and can be interpreted as topological defects.
  Gapped systems hosting fractons have been identified as certain exactly solvable models \cite{chamon,haah,vhf1,vhf2,bravyi,yoshida} and Chern-Simons gauge theories \cite{You2020BF,Ma2020CS,premabinav,pretkowitten}. They display topological ground state degeneracy accompanied by a unique entanglement structure. Unfortunately all these models are rather complicated which obstructs either their experimental realizations or even a detailed numerical study. In order to remedy this we construct a Bose-Hubbard type Hamiltonian in two dimensions with nearest neighbor interactions that is feasible to the numerical analysis and has a potential of being engineered and experimentally studied in recent state-of-the-art quantum simulator platforms with big time crystals \cite{giergiel2021} (for reviews see Ref.~\cite{Sacha2017rev,guo2020,SachaTC2020}).

The immobility of fracton excitations can be encapsulated in a generalized set of global conservation laws, that preserve various multipole moments of the charge density in addition to the total charge of the system. These conservation laws arise as a consequence of subsystem symmetry of the Hamiltonian. The question that we want to address is how to realize extended symmetries in a realistic Bose system with two-body interactions between the bosonic constituents taken into account? The standard Bose-Hubbard Hamiltonian is invariant under a global $U(1)$ symmetry. This means that a constant shift of the phase leave the theory invariant. As shown in Ref.~{\cite{Gromov_2019} a natural generalization of this symmetry to account for higher moment conservation is to employ the so-called space-dependent polynomial shifts \cite{Pretko2017spinliquid,Pretko2017spinliquid2}. The shift symmetries are broken by the usual hopping terms, which is natural since the symmetries encode the restricted mobility of excitations. Hence, the inclusion of constraints on the movement of particles in the model requires hopping terms that are engineered to account for fracton mobility constraints. One example of such terms is the ring-exchange term, which has been extensively studied in a context of exciton Bose liquids \cite{Paramekanti2002,Tay_2011,PhysRevB.75.104428,PhysRevB.67.134427} and more recently in the context of fractons \cite{You2020BF,You:2021tmm}. However, due to the failure of the mean-field treatment, the ring-exchange models are rather difficult to handle numerically, especially in the frustrated regime where quantum Monte  Carlo algorithms cannot be applied  \cite{Tay2011,Huerga2014}.

On the other hand, in this work we consider an extended Bose-Hubbard Hamiltonian at half filling with ring-exchange interactions but on $1\times2$ and $2\times1$ plaquettes in a strongly repulsive limit, where the low energy physics can be efficiently studied due to energetic constraints. This model provides a complementary realization of gapped fractons. Furthermore, as we argue below, the defect dynamics in this model resembles the constrained motion of defects seen for example in elasticity \cite{Cvetkovic_2006}. While the ground state of the system realizes a checkerboard charge density wave (CDW), by means of analytical and numerical analysis we find that the first excited band consists of fracton excitations (lineons) with the restricted mobility to one dimensional columns (or rows) of the two dimensional lattice. At the same time, we show that the second excited band hosts both completely immobile excitations and two joint excitations which can move freely on the entire lattice. In the later parts we investigate the robustness of our findings and concentrate on the experimental realization of the model. 


\section{The model}

\begin{figure}[t]
		\centering
		\includegraphics[width=0.75\columnwidth]{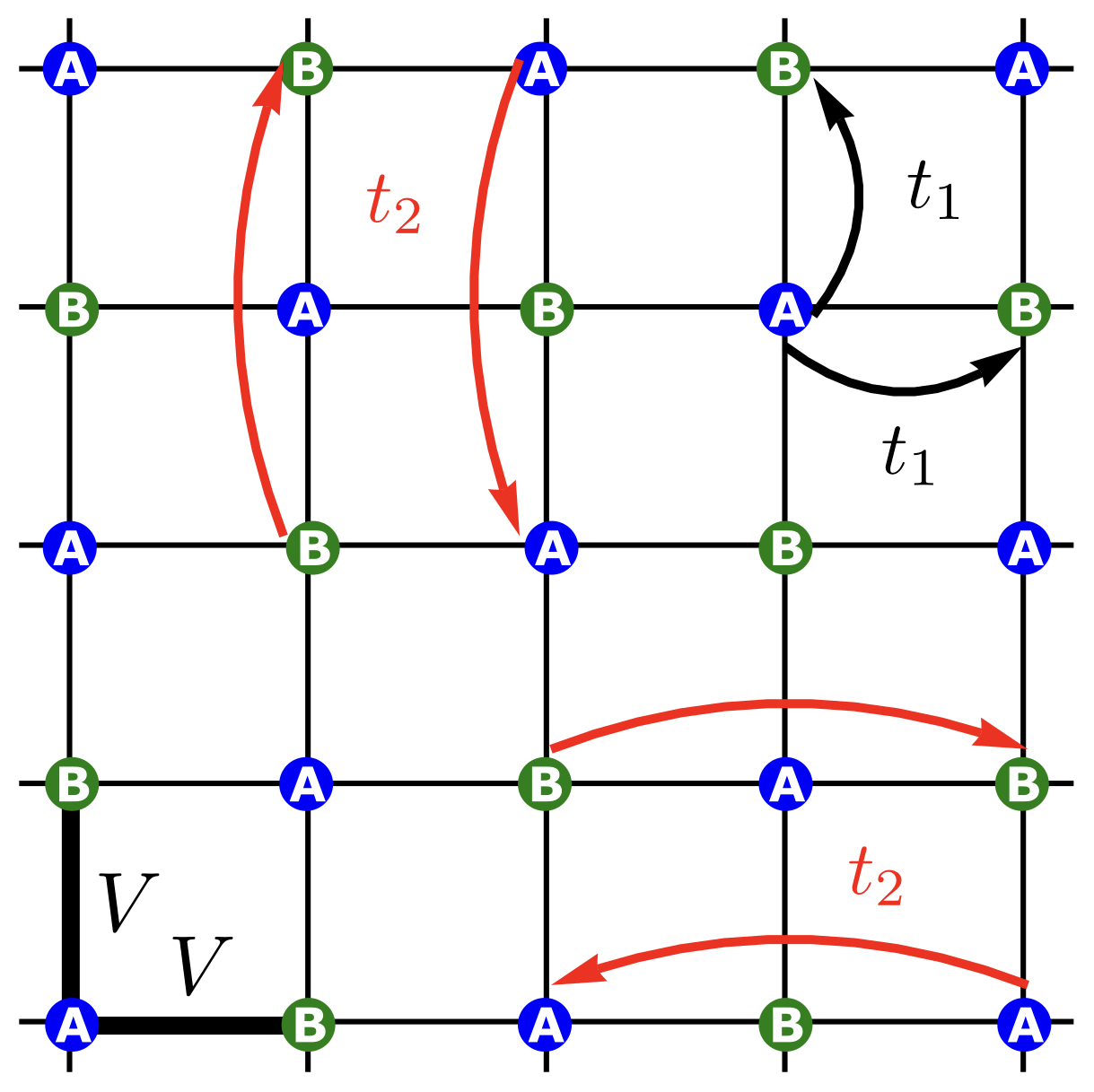}
		\caption{An illustration of all possible hopping and interaction processes in the model on a square 2D lattice: standard nearest neighbor hopping $t_1$,  ring-exchange terms $t_2$ and the nearest neighbor interaction $V$. 
		 The ring-exchange interaction we are considering consists of a simultaneous tunneling of two particles to the second neighbors along  $1\times 2$ and $2\times 1$ plaquettes. Site labels A and B denote two bipartite sublattices of a square lattice. Throughout the article we are considering strongly repulsive regime at half filling where, in the ground state, all the particles fill completely only one sublattice realizing the checkerboard charge density wave (CDW) order. The focus of our analysis is put on the lowest lying excited energy bands consisting of fracton excitations.}
		\label{hopping}
	\end{figure}
	
Bose-Hubbard models are simple but powerful models to investigate quantum phases and many-body dynamics of collective excitations 
\cite{Jaksch2005,sachdev2011quantum,Dutta2015}. 
Therefore Bose-Hubbard incarnation of fracton quasiparticles is desirable to understand the dynamics of these excitations. In order to achieve this let us consider a tight binding hard-core boson Hamiltonian with repulsive interactions on a two dimensional (2D) square lattice with periodic boundary conditions,
\begin{eqnarray} 
\label{modelmodel}
    \hat{H}   =   -t_1 \sum_{\langle  i j   \rangle }         \hat{b}^{\dagger}_i  \hat{b}_j     +  \frac{V}{2}  \sum_{\langle  i j  \rangle }       \hat{n}_i  \hat{n}_j  - t_2   \sum_{[  i j k l ] }   \hat{b}^{\dagger}_i  \hat{b}^{\dagger}_j  \hat{b}_k   \hat{b}_l , 
\end{eqnarray}
with $\hat{b}_i$ ($\hat{b}^{\dagger}_i$) denoting the standard anihilation (creation) operator of a bosonic particle localized on the $i$-th site of the lattice, $\hat{n}_i = \hat{b}^{\dagger}_i \hat{b}_i$ is the particle number operator. 
The first two terms of the Hamiltonian~\eqref{modelmodel} describe a familiar hard-core boson extended Bose-Hubbard model \cite{Dutta2015}  with $t_1$ and $V$ being the tunneling amplitude and the nearest neighbor interaction strength respectively. The symbol $\langle  i j   \rangle $ indicate the sum over nearest neighbors on the lattice.
The last term contains simultaneous tunneling processes of two particles, dubbed the ring-exchange interaction  \cite{Paramekanti2002,Sandvik2002,Rousseau2004,subsys2}, with a hopping amplitude~$t_2$. The ring-exchange interaction we are considering is a combination of two second neighbor hoppings along $1\times2$ and $2\times1$ plaquettes which preserves the center of mass of two particles and the number of particles in bipartite sublattices [cf. Fig.~\ref{hopping} for an illustration].  The symbol $[ijkl]$  denotes  the summation over all possible plaquettes. Throughout this article we will be considering strongly repulsive limit, i.e. $V\gg t_1,t_2$, where the ground state realizes the checkerboard charge density wave (CDW) order. A checkerboard CDW is a single Fock state where all of the particles occupy one of the two bipartite sublattices.
While in the next section we focus on the symmetries and conservation laws of this model which are necessary for the appearance of fractons, let us stress that the underlying CDW order stabilizes the fracton excitations we find in the lowest lying excited energy bands.

\section{Symmetries of the model}\label{sec:symmetry}

In this section we show that the extended Bose-Hubbard Hamiltonian with the ring-exchange term \eqref{modelmodel} possess certain symmetries which enable the model to host fracton excitations. Our analysis will be phrased in the language of multipole algebras constructed in Ref.~\cite{GromovLucas2020} (see also Ref.~\cite{Griffin:2014bta}). The basic ingredient of this construction is the dipole moment conservation, which follows from the so-called polynomial shift symmetries. These shift symmetries act on a scalar field $\varphi$ that in our case corresponds to the phase associated with creation and annihilation operators. Before we proceed let us recall basic facts about this construction. We consider a low energy dynamics of the phase field given by the action $S[\varphi]$. We assume that the action is invariant under the following transformation
\begin{equation}
\label{eq:Polyshift}
\delta \varphi = \lambda_{\alpha} \mathcal{P}^\alpha(\vec x) \,,
\end{equation}
where $\lambda_\alpha$ is a symmetry parameter and $\mathcal{P}^\alpha(\vec x)$ is a polynomial. It is enough for our purposes to consider only homogenous polynomials. Furthermore we assume that our theory is invariant under global shifts with a corresponding charge density $\rho(\vec x)$. It follows from the Noether theorem that charges
\begin{equation}
\label{eq:Charges2}
 Q^{(\alpha)} = \int  d^2 x \, \mathcal{P}^\alpha(\vec x) \rho(\vec x)\,.
\end{equation}
are conserved. Since the polynomials $\mathcal{P}^\alpha(\vec x)$ are homogenous the symmetry leads to a conservation of proper multipole moments. The main ingredient of this abstract construction is the conservation of the monopole and dipole charges.
\begin{subequations}
\begin{align}
    & \frac{d Q}{dt} =\frac{d}{dt} \int d^2 x \, \rho = 0, \\
    & \frac{d Q_i }{dt}  = \frac{d}{dt} \int d^2 x \, x_i \rho = 0,
\end{align}
\end{subequations}
where $x_i\in \{x,y\}$. In addition, if we assume that the system is translationally invariant, the momentum is also conserved
\begin{equation}
    \frac{d P_i}{d t} = \frac{d}{dt} \int d^2 x \, p_i = 0.
\end{equation}
In order to see the monopole and dipole conservation on a lattice, we first point out that a 2D theory has to be invariant under the following symmetry transformation
\begin{equation} \label{eq:Symmetric}
\delta \varphi = \lambda_0+\lambda_1 x +\lambda_2 y,
\end{equation}
with $\lambda_{\alpha=1,2}$ being symmetry parameters as in Eq.~\eqref{eq:Polyshift}.
Therefore, we need to check if the Bose-Hubbard model respects these symmetries. The interaction terms depend on the densities and the symmetry transformation in Eq.~\eqref{eq:Symmetric} follows immediately. We therefore only need to check the symmetry transformations of the hopping term. Of course the usual hopping implies full mobility and as such breaks the polynomial shift symmetry. However, the ring-exchange term preserves it. This can be seen explicitly by making a phase shift of the form
\begin{align}\label{symmetry}
    \varphi \rightarrow \varphi  + \zeta (x) + \chi (y) ,
\end{align}
which is a specific local U(1) phase transformation, that changes along rows or columns \cite{GromovLucas2020,PhysRevB.75.104428}, more general than the polynomial shift symmetry of Eq.~\eqref{eq:Symmetric}. The non-local ring-exchange term is invariant under the above phase transformation, which can be shown explicitly
\begin{align}
\begin{split}
  & \hat{b}^{\dagger}_i \hat{b}^{\dagger}_j \hat{b}_k \hat{b}_l \rightarrow  \hat{b}^{\dagger}_i \hat{b}^{\dagger}_j \hat{b}_k \hat{b}_l   \cdot e^{- i \zeta (x_1) -  i  \chi (y_1)} e^{- i \zeta (x_2) - i \chi (y_2)}  \\
  & \times e^{ i \zeta (x_2) + i \chi (y_1)} e^{ i \zeta (x_1) + i \chi (y_2)}   =    \hat{b}^{\dagger}_i \hat{b}^{\dagger}_j \hat{b}_k \hat{b}_l,
\end{split}
\end{align}
where we have used the fact that the ring-exchange interactions preserve the center of mass of the tunneling pair, cf. Fig.~\ref{hopping}. Note that our symmetry transformation in Eq.~\eqref{symmetry} implies a conservation of additional moments. Similar phenomenon occurs in other models as well. For example, the traceless scalar theory \cite{Pretko2017spinliquid2}, in addition to the conserved charges $Q,Q^i$, also conserves the moment $Q^{(2)}=\int d^2 x ||\mathbf x||^2 \rho$ and elasticity conserves one component of the quadrupole moment \cite{PretkoSolid2018}. The dipole moment conservation implies that an isolated charge cannot move. However, if additional moments are preserved it may put additional constraints on the movement of excitations. Conservation of the quadrupole moment implies that a dipole can only move in the direction perpendicular to its vector charge. In two dimensions such a movement is along a line and thus the quasiparticles giving rise to the dipole charge are called lineons.

In fact, the extended Bose-Hubbard model with any number of ring-exchange interaction terms regardless of their plaquette sizes $n_1\times n_2$ conserves a component of every $n$-th higher-moment of a given Cartesian coordinate
\begin{equation}
\frac{d Q_{i \ldots i}^{(n)}}{dt}  =  \frac{d}{dt}\int  d^2 x  (x_i)^n \, \rho (\vec x)  = 0,
\end{equation}
where $n\in \mathbb{N}_0$. One can actually choose for $\zeta(x)$ and $\chi(y)$ a family of functions which form a basis for any integrable functions of $x$ or $y$ over their respective domains. Completeness of the basis states allows one to restate this fact as conservation of marginal distributions of charge: 
\begin{subequations}
\begin{align}\label{marginals1}
f_1(x) & = \int d y \rho (\vec x), \\
f_2(y) & = \int  d x \rho (\vec x), \label{marginals2}
\end{align}
\end{subequations}
where functions $f_1$ and $f_2$ are conserved under ring-exchange interactions. It turns out the properties of fracton excitations can be easily understood by looking at the marginal distributions only, which we will elaborate on in the next section.

So far we have not used the fact the ring-exchange interactions in our model take place along rectangular $1\times2$ and $2\times1$ plaquettes. Such a plaquette choice is motivated by the presence of an additional symmetry which implies the conservation of the number of particles in each bipartite sublattice  
\be
[N_A,H]=[N_B,H]=0. 
\ee
Therefore, as long as the nearest neighbor interaction strength $V$ is a dominant energy scale, the energy spectrum is divided into isolated bands composed of Fock states with the same interaction energy. While the ground state is a simple CDW state, its first two excited bands consists of states with interaction energy $3V$ and $4V$ that differ from the CDW state by a single or double dislocations [cf. Sec.~\ref{sec:excitations} and Fig.~\ref{fracton}].  \ak{
Let us also stress that although any system with the ring-exchange and  the density-density interactions obeys subsystem symmetries leading to the multipole moments conservation, not all of such processes will lead to the mobility restrictions of localized excitations.  In our case,  due to the conservation of the number of particles in each bipartite sublattice, the ring-exchange term connects configurations with the same nearest neighbor interaction energy. As such, it constitutes a minimal Bose-Hubbard model with the ring-exchange interactions hosting single fractonic excitations. }
Due to the conservation laws discussed in this section, \ak{which are exact for $t_1=0$ and approximate for $t_1\ne0$, } the Hilbert space of dislocations is fragmented into disjoint sectors. As we show in the next section the latter enforces the mobility restrictions of the excitations. \ak{We note that this fragmentation is distinct from the dynamical Hilbert space shattering \cite{tiltref3,tiltref5,Iaconis2021,tiltref6} where fragmentation happens for fixed symmetry sectors due to dynamical constraints, which are responsible,  \rub{among others}, for the subdiffusion in the tilted Fermi-Hubbard models \cite{tiltref1,tiltref2}.}

%



\begin{figure}[h!t]
		\centering
		\includegraphics[width=0.95\columnwidth]{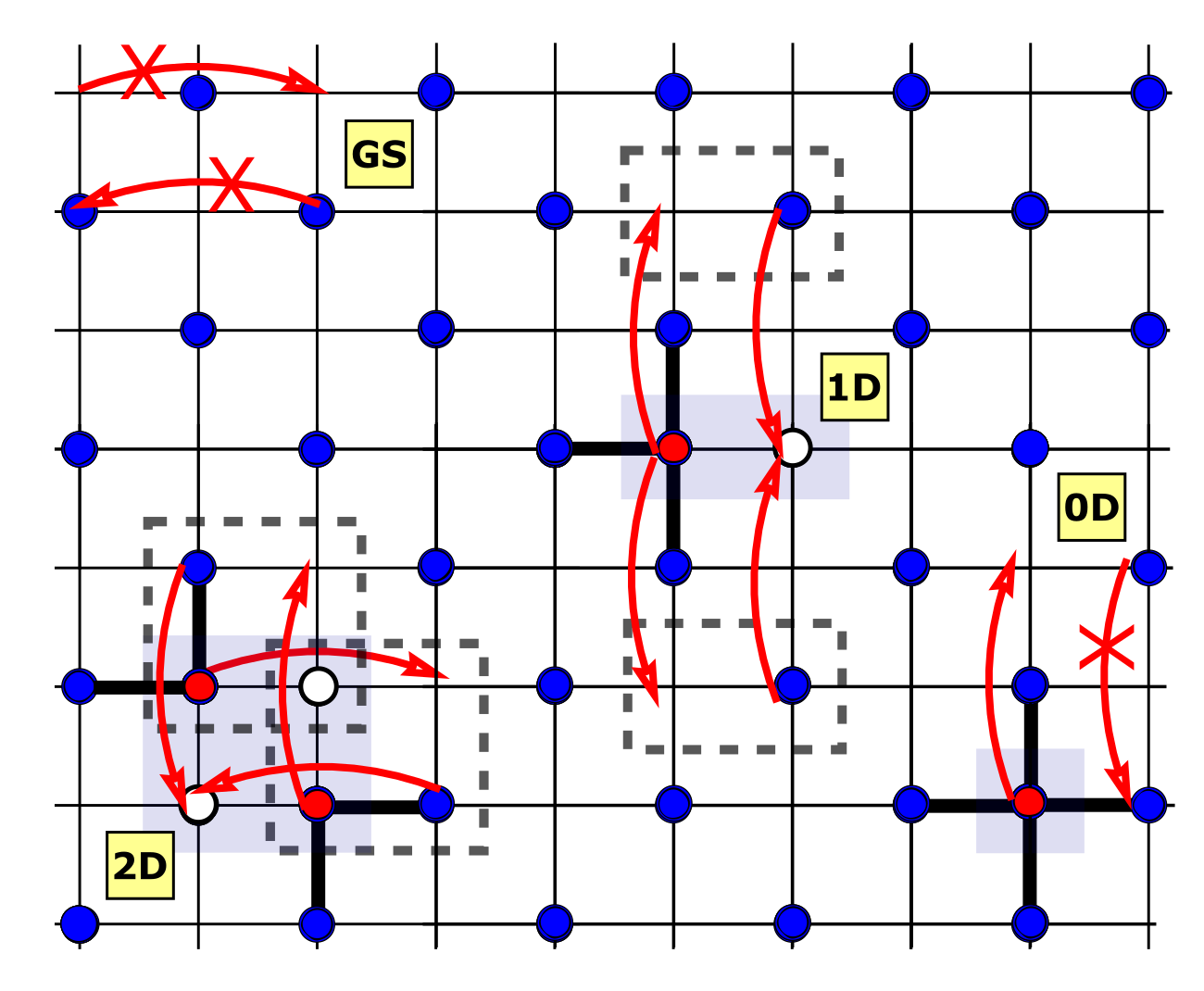}
		\caption{Schematic view of the CDW dislocation movement due to ring-exchange interactions. White circles show particles missing from the checkerboard pattern and red circles show particles present in a sublattice, which is empty in the ground state. A red cross indicates a prohibited hopping. When the system is in the ground state (GS) there are no dislocations and no hopping can take place. For non-local dislocations, hopping is still not possible, and hence the motion of such excitations is zero-dimensional (0D). For a single local dislocation one has one-dimensional (1D) motion and for a double local dislocation one has two-dimensional (2D) motion.} 
		\label{fracton}
	\end{figure}

\section{Fracton excitations}\label{sec:excitations}
 Before we turn to the description of fracton excitations and their dynamics we start this section by looking at lowest energy excitations of the model   in a strongly interacting regime $V\gg t_1,t_2$ we consider in this work. These lowest energy excitations can be constructed by looking at  single particle dislocations in the CDW ground state. A single dislocation state is created when one creates a particle on an empty site in the CDW state and also annihilates a particle on the occupied sublattice to preserve half filling. The space of lowest energy excitations consists of single local particle dislocations
\begin{subequations} \label{statesdefect}
\begin{align}\label{dipolex1}
\left(D^{x}_{ij}\right)^\dagger
|\textrm{CDW}\rangle
 &=\left(\hat b_{i+1,j}^\dagger \hat b_{i,j} 
 +\hat b_{i,j}^\dagger \hat b_{i+1,j}
 \right)
 |\textrm{CDW}\rangle,  \\
\left(D^{y}_{ij}\right)^\dagger
|\textrm{CDW}\rangle
 & = \left(\hat b_{i,j+1}^\dagger \hat b_{i,j} 
 +\hat b_{i,j}^\dagger \hat b_{i,j+1}
 \right)
 |\textrm{CDW}\rangle,  \label{dipolex2}
\end{align}
\end{subequations}
whose interaction energy is  $3V$. 
Note that the above definition is indifferent to the choice of a filled sublattice in the charge density wave.
If the dislocation is not local i.e. the two sites that deviate from the CDW state are not nearest neighbors, then such a state belongs to the second excited energy band with an eigenenergy of $4V$. The second excited energy band is completed with  states that compose of two neighboring local dislocations 
\begin{align}
|D^{(2)}_{ij}\rangle &=
\left(D^{x}_{ij}\right)^\dagger \left(D^{x}_{i,j+1}\right)^\dagger
|\textrm{CDW}\rangle  \nonumber \\
&=   \left(D^{y}_{ij}\right)^\dagger \left(D^{y}_{i+1,j}\right)^\dagger
|\textrm{CDW}\rangle.\label{double}
\end{align}
Other states, such as non-neighboring double dislocations, belong to higher energy bands and are beyond the scope of interest of this article. In the next parts we argue the single dislocation states correspond to either completelely  immobile zero dimensional fractons or  partially mobile lineons (one dimensional fractons),  whereas the two dislocation state $|D^{(2)}_{ij}\rangle$ can be interpreted as two bound lineon states which is allowed to move in both directions, see Fig.~\ref{fracton} for an illustration.  First, we support our arguments by constructing approximate eigenstates of the model in the  $t_1 \ll t_2$ limit and then we numerically study the robustness of fracton excitations using exact time propagation.

\begin{figure*}[tbh!]
\centering
\includegraphics[height=8cm]{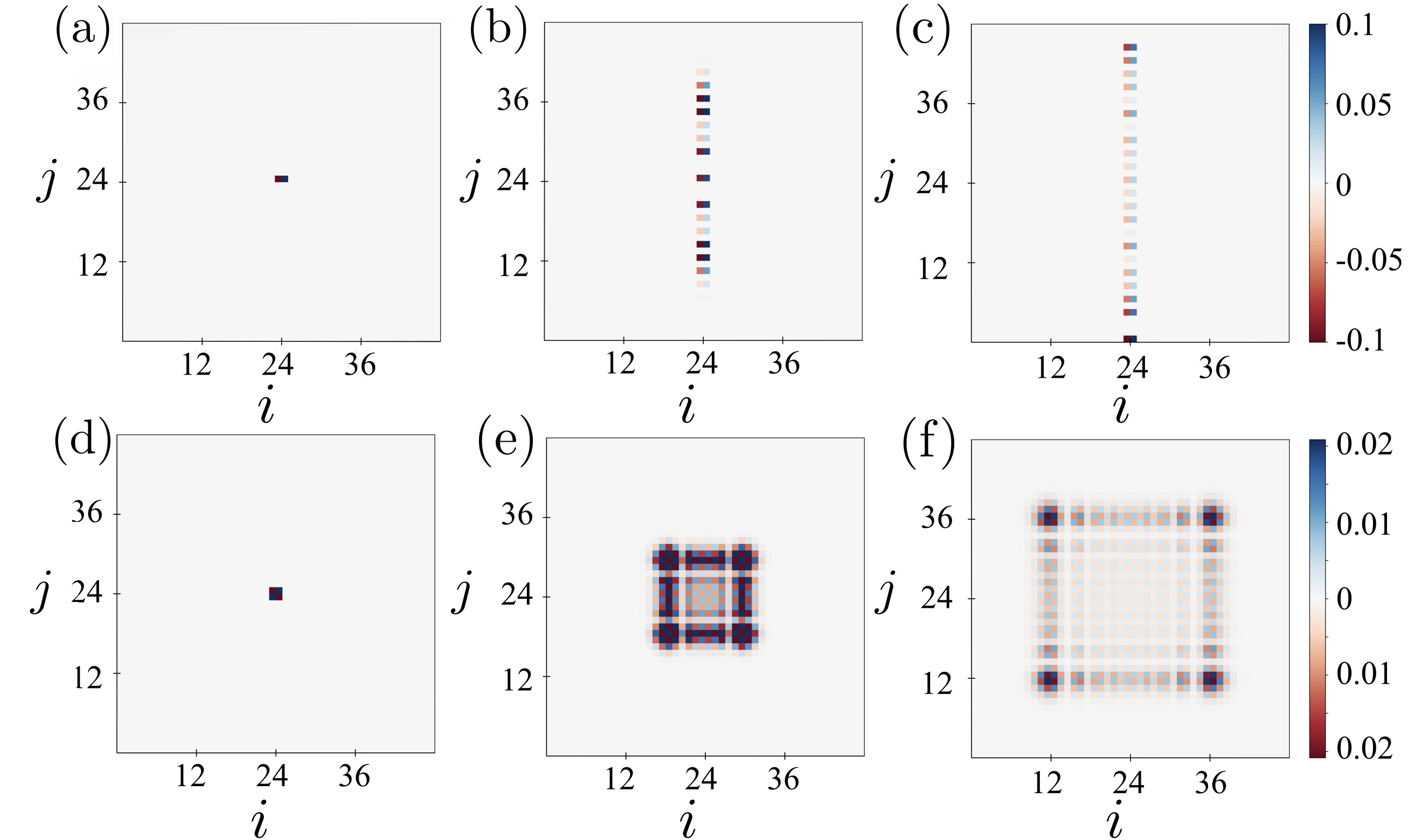}
\caption{Exact time propagation of elementary dislocations. While panels (a-c) show a propagation of a single local dislocation, panels (d-f) illustrate a motion of a bound double dislocation.  Each row is associated with time point $\tau=0$, $\tau=3.5/t_2$ and  $\tau=7/t_2$. The panels confirm the analytical arguments of the main text that a single local dislocation can propagate only in one dimension but two bound dislocation state propagates freely in two dimension.  The movement of the later is twice slower compared to a single dislocation state.
The lattice size is 48x48 sites, each site is represented by a single pixel. Positive probability (blue) shows particle density in empty lattice, as compared to the CDW ground state. Negative probability (red) is used to show hole density, as compared to the CDW ground state. The color scale is cut at low values to be consistent along all images in a column.  Model parameters are $t_1=0$, $t_2=0.01V$. }
\label{TimePropDisl}
\end{figure*}

\begin{figure}[t!]
		\centering
		\includegraphics[width=1.\columnwidth]{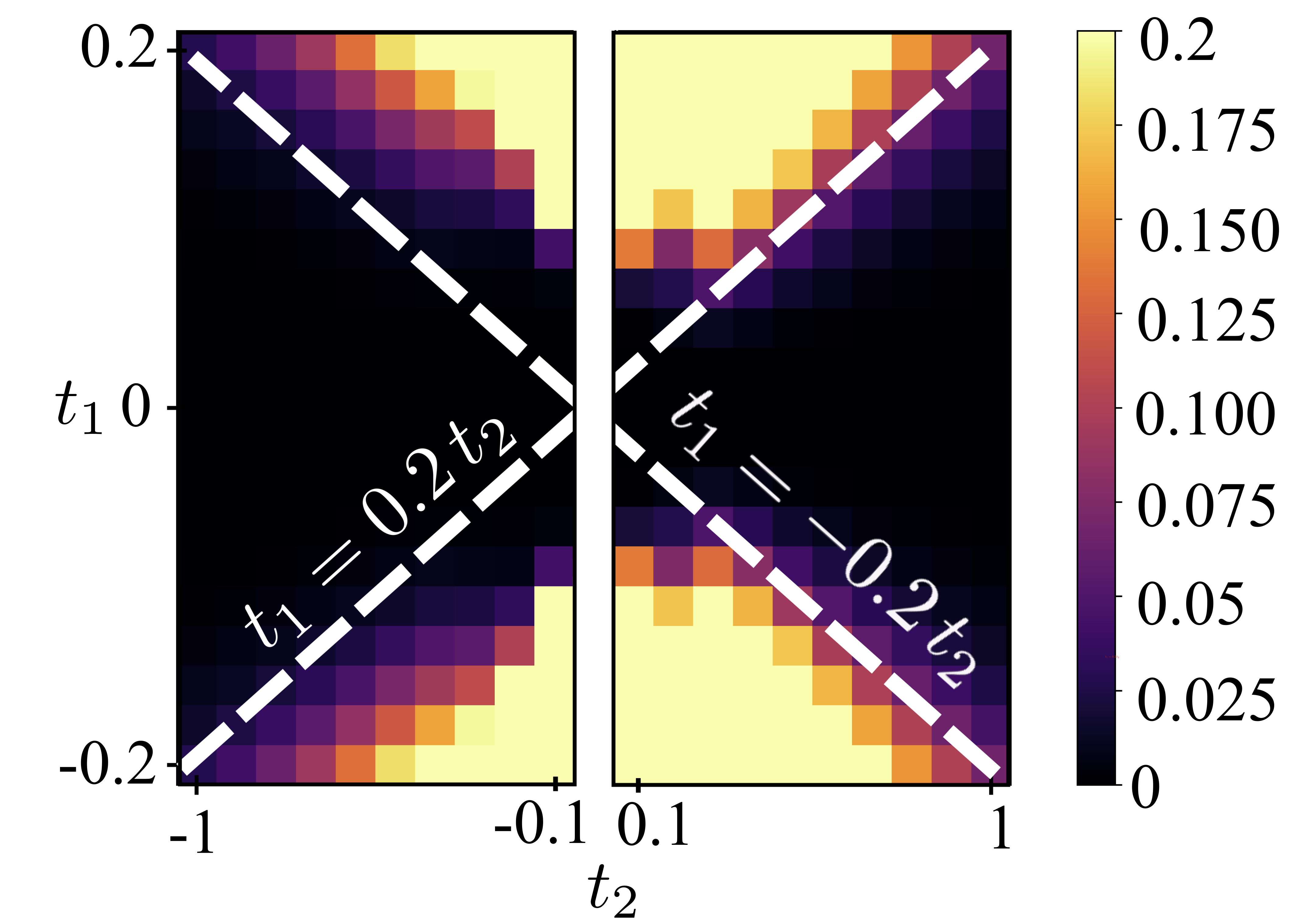}
		\caption{Stability of the fractonic behavior of a single initial dislocation as seen in the top-left panel of Fig. \ref{TimePropDisl}. The color scale represents \ak{root} sum of squared differences of initial and final marginal distributions. The plotted probability is the value achieved during time evolution on $20\times 20$ lattice for time $\tau=2/t_2$. $V=1$ is assumed.}
		\label{exactexo}
	\end{figure}

\begin{figure*}[tbh!]
\centering
\includegraphics[height=4.5cm]{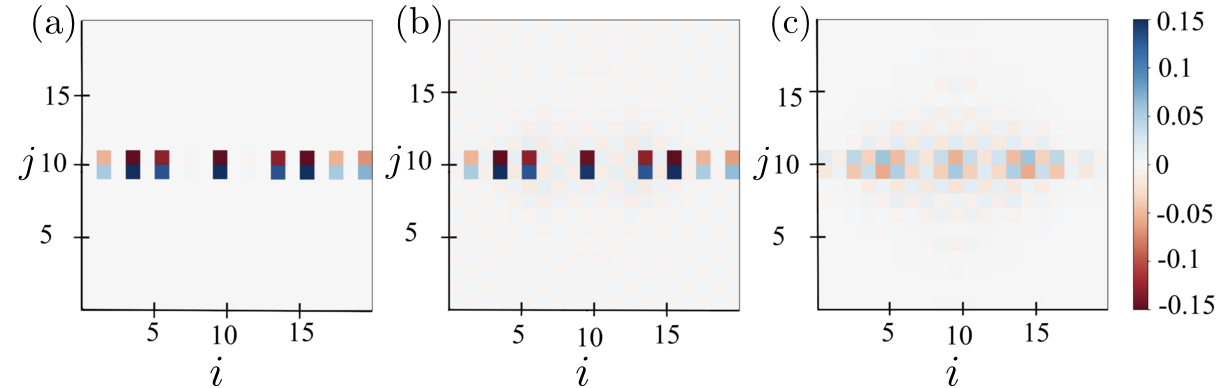}
\caption{Exact time propagation for a single dislocation at time $\tau = 4 / t_2 $ for $t_2 =1  V$ and (a) $t_1= 0$, for (b) $t_1 = 0.2 V$ and for (c) $t_2 = 0.1 V$ and $t_1 = 0.2 V $. The lattice size is 20x20 sites, each site is represented by a single pixel. }

\label{TimePropDis11l}
\end{figure*}

\subsection{Fracton eigenstates}
\label{spectrum}

Let us first look at the eigenstates of the Hamiltonian when normal hopping is set to zero, i.e. $t_1 = 0$. First of all, since the ring-exchange interaction is a combination of two simultaneous  local particle hoppings, the non-local dislocation state is a completely immobile eigenstate of the model with the eigenenergy equal to the interaction energy $4V$ and corresponds to two isolated fractons, with one being a particle, and the other one a hole [cf. Fig.~\ref{fracton}]. The first mobile excitation one can have is a fracton dipole which can be constructed from single local dislocation states [cf. Eq.~\eqref{dipolex1}-\eqref{dipolex2}]. The fracton dipoles can  only move orthogonally to the dipole moment \cite{You_emergent_2020,subsys2,Pretko2017spinliquid2}.  In two dimensions that means the state is only allowed to move along a line and thus these states are also called lineons [cf. Fig.~\ref{fracton}]. The momentum state of fracton dipole with momentum orthogonal to the dipole moment pointing in the $y$-direction reads  \cite{sous2019fractons} 
\begin{align}
    |  k_{x} , m        \rangle   = \frac{1}{\sqrt{L} }  \sum_{n}     e^{2 i a  n k_{x}}  
    \left(D^{y}_{2n,m}\right)^\dagger
|\textrm{CDW}\rangle ,  \label{excitations}
\end{align}
where $L=L_{x}=L_{y}$ is the linear lattice size, $a$ is the lattice constant. The state $ |  k_{x} , m \rangle $ is an eigenstate of the Hamiltonian, which we can check explicitly, i.e.,
\begin{align}
   \frac{1}{2}  \sum_{ \langle  k l    \rangle  } \hat{n}_k  \hat{n}_l   |  k_{x} , m   \rangle  &  =     3 \, |  k_{x} , m  \rangle   ,   \\ 
       \sum_{ [  k l m n  ] }     \hat{b}^{\dagger}_k \hat{b}^{\dagger}_l  \hat{b}_m  \hat{b}_n     |  k_{x} , m   \rangle   &   =
    2   \cos(2 k_x a  )   |  k_{x} , m   \rangle   .
\end{align}
It is easy to find out that the same results will be given by  an analogous momentum state  $ |  k_{y} , n \rangle $  for dipoles  pointing in the $x$-direction.   We thus conclude that the energy dispersion of the first excited band is 2L-fold degenerate and given by
\begin{align}
E^{(1)}(k_i,j)  =     3 V  -     2  t_2  \cos(2 k_{i} a   ).
\end{align}
This restricted mobility of lineons can be also understood from the marginal distributions. The charge density state has constant (featureless) marginal distributions $f_1(x)$ and $f_2(y)$, Eq.~\eqref{marginals1}-\eqref{marginals2}. The $x$ dipole is generated from CDW by moving particle in direction $x$, which leaves a hole and creates a peak in $f_1(x)$ distribution and leaves $f_2(y)$ featureless. This means that there can be no movement in direction $x$ due to ring-exchange interaction. 

In a similar way  we can construct the second kind of mobile excitations in the lowest energy sector as a momentum state in a two bound dislocations subspace [cf. Eq~\eqref{double}],
\be
 | k_x,k_y  \rangle = \frac{1}{L} \sum_{m,n}     e^{i (k_x m+k_y n)a}    \label{4vstate}
|D^{(2)}_{ij}\rangle, 
\ee
which is  an approximate eigenstate of the  model with the  eigenenergy
\begin{align}
E^{(2)} (k_x,k_y) \approx     4 V  -     2  t_2  \left[\cos( k_x a  )+  \cos( k_y a  )\right].
\end{align}
The full mobility of this state can be explained by the fact that its marginal distributions are featureless and look the same as for ground state. The latter is understandable since, due to the indistinguishability of bosons, a state $|D^{(2)}_{ij}\rangle$  can be viewed as two  $x$ or two $y$ dipoles [cf. Eq~\eqref{double}].

If the two dislocations were far apart, then its interaction energy would be equal to $6V$, twice the energy of a single lineon. However, two dislocations close to each other form a bound state with an interaction energy $4V$. The momentum state Eq.~\eqref{4vstate} is only approximately the eigenstate because in principle two dislocations can dissociate when either moves in the opposite direction, but this would require a surplus energy to break a bound state.  Only processes that keep the bound state together do not change its interaction energy. Each ring-exchange interaction moves the center of mass of a bound state by one lattice site, an hence  the state propagates with twice smaller group velocity $v= \mathrm{max} [\partial_{ k_i}  E^{(2)} ] $.

\subsection{Time evolution of localized dislocations}

In this section we first confirm numerically the considerations of the previous section and then investigate the robustness of our findings by looking at the time propagation of dislocations with non-zero normal hopping on smaller lattices. 
The numerical time propagation is feasible when we are in the strongly interacting regime of parameters and can cut the Hilbert space using the  energy criterion.

\begin{figure*}[tbh!]
\centering
\includegraphics[height=13cm]{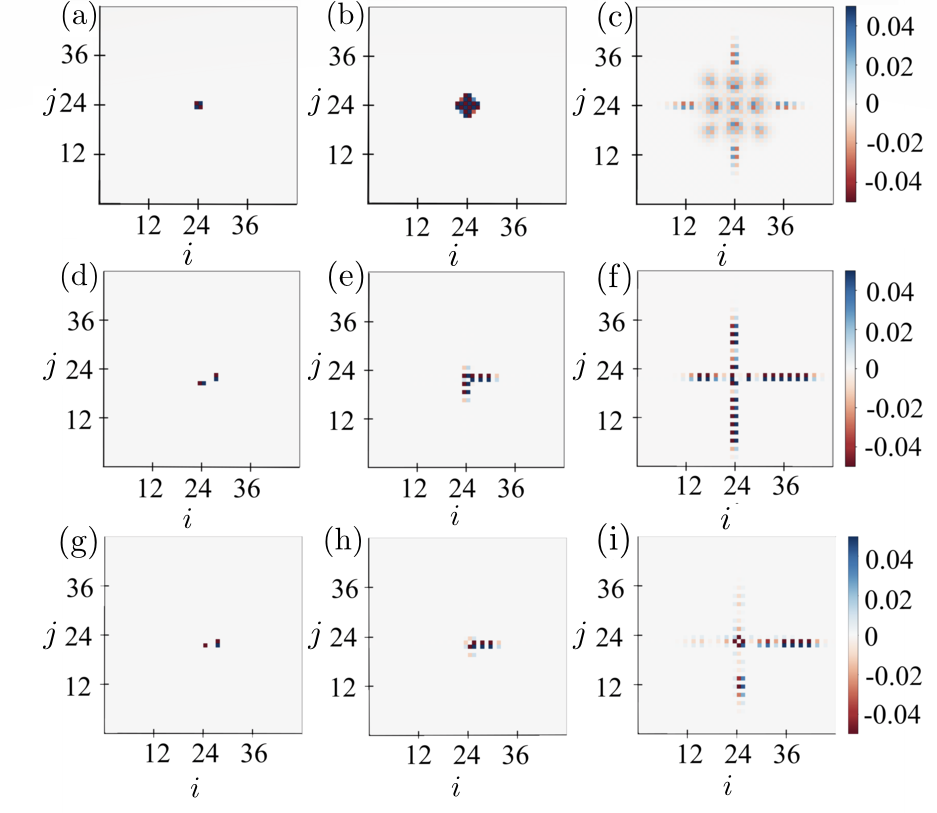}
\caption{\ak{Exact time evolution of few fractons for $t_1=0$. Columns present snapshot of the evolution for times $\tau = 0, 1, 7$. Panels (a-c) show the evolution of a double defect for $t_2 =0.5 V$ 
Here the value of $t_2$ is much larger than in panels (d-f) in Fig~\ref{interactionnnnn}, and because of this, one-dimensional defects, which have a larger interaction energy, appear out of the squared shape because of their larger group velocity. Panels \rub{(d-f)} show the evolution of two orthogonal separated single defects for $t_2 =0.5 V$. We see that these defects move straight through eachother and no two-dimensional mobility arises \ak{, but due to their interaction the time evolution is asymmetric as the propagation is delayed after the collision. Panels (g-i)}  show the scattering of a lineon on a \rub{zero-}dimensional immobile fracton.} } 
\label{interactionnnnn}
\end{figure*}

The Hilbert space dimension for $20\times 20$ lattice is then about $2\cdot 10^6$ for repulsive energies $E_R\le 5V$, which is perfectly fine for vector exponentiation using Krylov techniques \cite{moler2003nineteen}. We first discuss the ideal case at $t_1=0$ where the symmetries are perfectly satisfied. In Fig.~\ref{TimePropDisl} a single and double defect is initialized and time-evolution is studied. Panels (a-c) show that a localised local dislocation can propagate only in a direction orthogonal to its dipole moment. On the other hand, as shown in panels (d-f), two close dislocations form a bound state which does not have any mobility restrictions. 

Now we consider the case of a finite $t_1 $. For $t_1 \neq 0 $, the defects will eventually dissolve, because the normal hopping operator allows to transition to states that have a different number of defects.
\ak{However,  our considerations concern states that only locally deviate from the ground state. Although the local dislocations will in principle decay for a finite $V/t_1$, due the Lieb-Robinson \cite{Lieb1972} bound the speed of the information propagation will be always finite ($\sim 2|t_1|$), and therefore a state  deviating locally from a ground state will never quantum thermalize for an infinite system. For this reason, the stability of fractonic excitations for $t_2\ne 0$  is based on the stability of the charge density. A zero dimensional (0D) immobile excitation is an exact eigenstate due the hardcore boson constraint and therefore it does not thermalize at all. Of course, the lack of quantum thermalization does not imply that the 1D fractonic behavior will be always present in all parameter regimes. To guarantee the latter, the lineon group velocity $(4|t_2|)$ must an order of magnitude greater than the Lieb-Robinson velocity, i.e. $|t_1| \ll 2|t_2|$. 
} To show the strength of this effect, in Fig.~\ref{exactexo} we consider the degree to which a single defect signature disappears from the marginal distribution after some time $\tau = 2/t_1$. We see that for larger values of $t_1$, the destablizing effect of the normal hopping operator $t_1$ starts to become noticeable past lines of $t_1 \approx \pm  0.2 t_2$, \ak{which agrees with our theoretical preditions.} Note that Fig.~\ref{exactexo} is symmetric for $t_1 \rightarrow - t_1$, which is because the normal hopping operator plays the same role of creating and annihilating defects regardless of the sign of $t_1$. \ak{There is no such symmetry for $t_2 \rightarrow - t_2$, \rub{because for finite $t_1$ and negative $t_2$ there is frustration which is not the case for positive $t_2$}. Moreover}, in Fig.~\ref{TimePropDis11l}, we show the evolution of \ak{single dislocations} after a time $\tau$ for some specific cases values of $t_1$ and $t_2$. It is clear from comparing the ideal case of panel (a) to panel (b) that for a small value of $t_1$ the mobility restrictions effectively remain, whereas for large $t_1$ compared to $t_2$, as shown in (c), the behavior \ak{starts to become} two-dimensional, \ak{but for a finite evolution time it is still possible to distinguish a one dimensional density profile.} 

\ak{Finally, we stress that our approach allows for the study of more complex phenomena related to the evolution of fractonic defects, such as few and many fracton interaction dynamics with and without the presence of disorder. Although a detailed analysis of interaction phenomena goes beyond the scope of this paper, in Fig~\ref{interactionnnnn}  we present some basic results: panels (a-c) show the decays of a two dislocation state into two orthogonal lineons,  panels (d-f) \rub{show} lineon collision and panels (g-i) \rub{show} a scattering of lineon on an immobile (0D) fracton.}

\section{Experimental proposal}\label{sec:exp}

A pleasant feature of the model \eqref{modelmodel} in a strongly repulsive regime at half filling is that its low-energy physics can be studied analytically (for $t_1=0$) and efficiently simulated numerically (for $t_1\ne0$) which is due to the energetic cut-off of the basis states. Nevertheless other parameter regimes still present a challenge for a numerical analysis. On the other hand, for a few decades now, quantum simulators have provided a fertile ground for the study of many body physics. Recently it was also realized that condensed matter phenomena can appear in time crystals. Although most theoretical \cite{Wilczek2012,Sacha2015,Khemani16,ElseFTC,
Yao2017,Lazarides2017,
Russomanno2017,Ho2017,
Huang2017,Iemini2017,Wang2017,
Zeng2017,Surace2018,
Mizuta2018,Giergiel2018a,
Kosior2018,Kosior2018a,
Pizzi2019a,liang2018floquet,
Bomantara2018,Fan2019,Kozin2019,
Matus2019,Pizzi2021,
SyrwidKosiorSacha2020,Syrwid2020,
Russomanno2020,giergiel2020creating,Wang2020,kuros2020} and experimental \cite{Zhang2017,Choi2017,Pal2018,Rovny2018,Autti2018,
Kreil2018,Rovny2018a, Smits2018,Liao2018,Autti2021} works on time crystals have focused on the period-doubled discrete time crystals, the upcoming gravitational bouncer experiment  \cite{giergiel2020creating} is expected to demonstrate {\it big time crystals}, where discrete time translation symmetry broken states $\phi_n(\tau)$, $n=1,\ldots, s$ evolve with a period $sT$, up to $s=100$ times longer than the periodicity of the drive $T$. The symmetry broken states $\phi_n(\tau)=\phi_{n+1}(\tau+T)$ are localized wavepackets which evolve periodically in the laboratory frame of reference along a classical trajectory of the particle in the gravitational field. As such,  they can be interpreted as temporal analogs of the familiar Wannier states localized on temporally equidistant points on the closed classical trajectory. Although the temporal Wannier states $\phi_n(\tau)$ break time translation symmetry of the periodic potential, they constitute a convenient basis to study many-body temporal lattice models. Once the field operator $\hat\psi$  describing particles in a periodically driven system is restricted to the states $\phi_n(\tau)$, the effective Floquet Hamiltonian $H_F=H(\tau)-i\partial_\tau$ describing the quasi-energy structure of the system takes the form of a Bose-Hubbard Hamiltonian
\bea\label{eff_ham}
 H_F&=&\frac{1}{T}\int_0^{T}d\tau\int dx dy\;\hat\psi^\dagger\left(H_0(\tau)+\frac{g_0}{2}\hat\psi^\dagger\hat\psi-i\partial_\tau\right)\hat\psi
\cr &\approx&t_1 \sum_{<i,j>}\hat b_{i}^\dagger b_j+\sum_{ijkl}U_{ijkl}\hat b^\dagger_i\hat b_j^\dagger\hat b_k\hat b_l,
\eea
where $H_0(\tau)=H_0(\tau+T)$ is a single particle's time $T$-periodic Hamiltonian, $\hat\psi\approx\sum_n \phi_{j}(\tau)\hat b_j$ is a field operator with bosonic operators $\hat b_j$ annihilating a particle occupying the $\phi_{j}$ state. The parameter $t_1$ is the tunneling amplitude between neighboring sites and $U_{ijkl}$ is the interaction strength of a two particle scattering process between sites $k,l$ and $i,j$. Note that the quasi-energies of a time periodic problem are unbounded and defined up to a shift by $\omega = 2\pi/T$. Here we assume that $\omega$ is large compared to other energy scales in a system so that an infinite Floquet matrix can be reduced to a single diagonal block  as in the effective Hamiltonian \eqref{eff_ham} \cite{Eckardt2017}.

The above description can be generalised to all motional degrees of freedom of a particle, which would result in multidimensional time crystals \cite{SachaTC2020}. In particular, a particle moving along closed periodic trajectories in two orthogonal spatial directions in a gravitational bouncer model is described by $\phi_{n,m}(x,y,\tau)=\psi^{(x)}_{n}(x,\tau)\psi_{m}^{(y)}(y,\tau)$,  where $\psi^{(x_i)}_{n}(x_i,\tau)=\psi_{n+1}^{(x_i)}(x_i,\tau+T)$.
Another relevant feature of the gravitational bouncer model is that the microscopic parameters of $H_F$ can be in principle manipulated experimentally by means of the Feshbach resonance \cite{Chin2010} which modifies atomic interactions and hence the coefficients $U_{ijkl}$.
 This gives unique possibilities to investigate the full phase diagrams of various many body lattice models experimentally. 

In a recent Letter \cite{giergiel2021} it was shown that the gravitational bouncer model with two periodically oscillating mirrors has properties of a two-dimensional time crystal where the geometry of a temporal lattice can be arbitrarily shaped. In particular, one example shows how to design a Lieb lattice with a well separated middle flat energy band where the dynamics is 
governed by a many-body Hamiltonian with long range density-density and long-distance ring-exchange interactions fulfilling fracton symmetries discusses in Section~\ref{sec:symmetry}. This physical system should host fractons, but possesses additional complications, because the Bose-Hubbard model analyzed in Ref.~\cite{giergiel2021} exists on a M\"obius strip and the ring-exchange term does not exhibit the full lattice translation symmetry. 

Here we overcome these difficulties by showing how to realize a minimally extended Bose-Hubbard model hosting fractons, as in Eq.~\eqref{modelmodel}. The experimental proposal relies on two basic steps:
\begin{enumerate}
\item Effectively switching off the nearest neighbor tunnelings $t_1$ by  building a deep  temporal lattice with well localized states on a 2D torus, where $t_1$ are negligible.
\item Selecting the relevant interactions by the Feshbach resonance.
\end{enumerate}
In order to achieve this we choose a symmetric drive of two orthogonal mirrors giving rise to a square $n\times n$ temporal lattice  [cf Fig.~\ref{exp}(a)]. In the first step only a subset of ring-exchange interactions are realised. However, one can restore the full translation symmetry by adiabatically moving lattice in the $x$ or $y$ direction [cf Fig.~\ref{exp}(b-c)]. After $n$ such steps, all of the ring-exchange tunnelings are realized and the time averaging reproduces exactly the Hamiltonian \eqref{modelmodel}, which we analyze in the main part of this paper. A schematic of the protocol is presented in Fig.~\ref{exp}(d). Note that the limit on the speed of the adiabatic shift of the lattice  is given by the energy separation of the lowest and first excited quasi-energy bands, which is the higher, the deeper is the lattice. This works in a favour of our experimental proposal, as the fractons are expected in parameter range $t_1\approx 0$, where the driving change can be the fastest.

\begin{figure}[!ht]
\centering
\includegraphics[width=1.0\columnwidth]{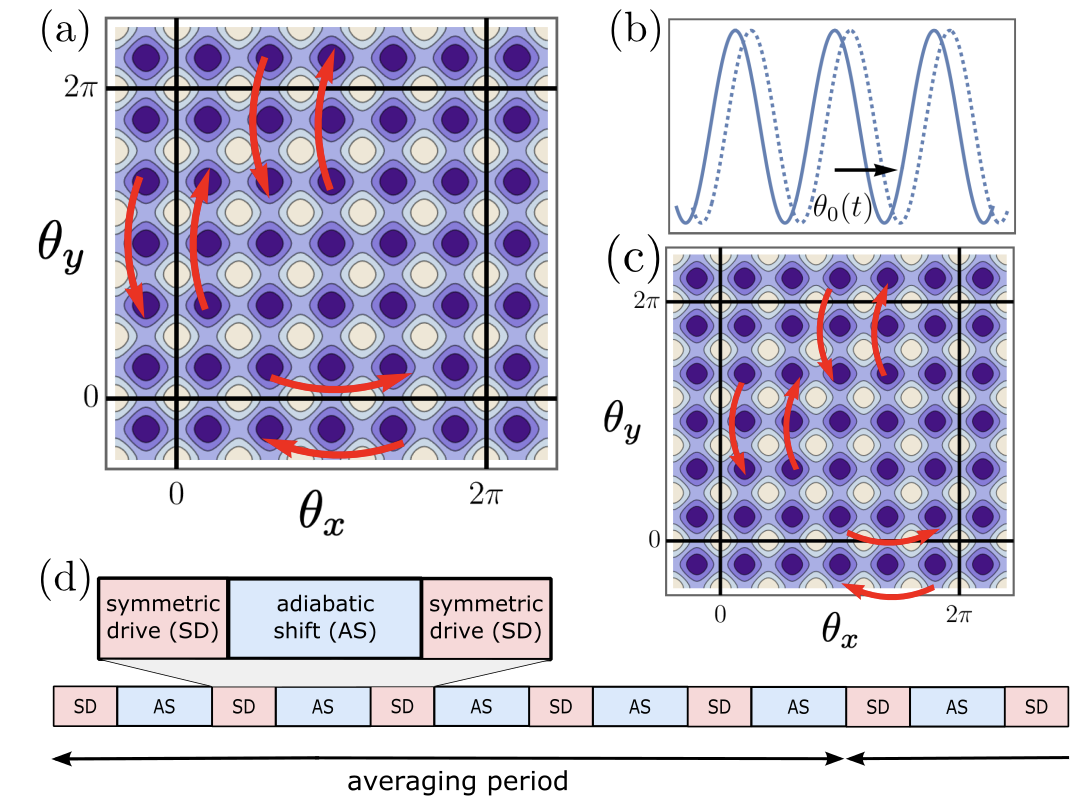}\\
\caption{
Schematic of the experimental protocol. Panels (a-c) illustrate subsequent building blocks of the protocol, which is summarized on panel (d). Note that for illustrative purposes each panel presents a relatively small $5\times 5$ lattice, but experimentally relevant sizes can go up to $100\times100$ sites [cf. Ref.~\cite{giergiel2020creating} for  accessible lattice sizes in 1D experiments]. Panel~(a) shows a 2D time crystal with $2\times 1$ and $1\times 2$ ring-exchange interactions. The axes $\theta_x$ and $\theta_y$ represent directions of the flat torus shown as black rectangle, with parts of it repeated to illustrate the boundary conditions. Such a lattice is achieved using drives: $\text{drive}_x(\tau)=\text{drive}_y(\tau)=\cos(5 \omega \tau)$. Note that in the first step only a subset of the ring-exchange interactions is realized and only a few of them are shown for illustrative purposes. 
Panel (b) shows the adiabatic movement step, which is introduced to restore the full translational symmetry of the effective Hamiltonian. 
This is achieved by introducing a slow adiabatic shift in one of the drives, e.g. $\text{drive}_x(\tau)=\cos[ 5 \omega \tau  +\theta_0(\tau)]$. This drive change is stopped after the lattice is moved by a single position to the right $\theta_0(\tau_{end})=2\pi$, which  also means that the $x$ drive made one more oscillation then the $y$ drive. After this step new set of ring-exchange interactions is realized [panel (c)]. After $5$ such steps, we return to original positions. Time averaged Bose-Hubbard Hamiltonian realizes exactly the model \eqref{modelmodel} described in this article.}
\label{exp}
\end{figure}
\twocolumngrid\

\section{Conclusions and perspectives}

 We have studied an extended Bose-Hubbard model with ring-exchange interactions in a strongly interacting limit at half filling. We have argued that this model has a ground state that corresponds to a charge density wave and hosts fracton excitations in the excited states, which appear due to mobility constraints arising from symmetries and conservation laws of the model. We have studied the dynamics of these excitations and  concluded that a single fractonic defect is either completely immobile or can propagate only in one dimension (lineon). On the other hand,  a bound state of two lineons is free to propagate in all directions. We have argued that these properties can be also understood in terms of conserved marginal charge distributions. We have also addressed the stability issue, showing that these states are stable against small symmetry breaking perturbations such as the usual nearest neighbor hopping term. Finally we have devised an experimental proposal to realize the fractonic extended Bose-Hubbard model in the upcoming platforms hosting big time crystals. As a result our construction provides a clear path towards understanding of many body dynamics with restricted mobility.

The most interesting future extension of this work involves the investigation of the full phase diagram of our model.
 Although the analysis of this paper has been restricted to the excited states of the charge density wave,  it is known that both a simple and extended Bose-Hubbard models can enter a superfluid regime \cite{Jaksch2005,sachdev2011quantum,Dutta2015} characterized by a spontaneously broken $U(1)$ symmetry and associated Goldstone modes.
The notion of a fractonic analogues of superfluids was recently introduced in \cite{peng1,peng2} (see also \cite{Argurio:2021opr}), using low-energy Hamiltonians defined in continuum.  It is therefore desirable to have a microscopic understanding of such phases and a detailed study of phases and phase transitions. Although arguments  presented in Ref. \cite{peng1,peng2} based on the Mermin-Wagner theorem indicate that $d=3$ is the lowest stable dimension at zero temperature, a phase with a quasi long-range order is possible in two spatial dimensions. This is similar to the ordinary Bose-Hubbard model, which at zero temperature can have a quasi long-range order for $d=1$ and true long-range order only for $d=2$ \cite{lewenstein2017ultracold}. However, the phase with a quasi long-range order will effectively be a superfluid phase for small system sizes. A detailed numerical study can shed light onto exotic superfluids which go beyond the mean field treatment.

\section*{Acknowledgments}
We acknowledge discussions with Arkadiusz Kuro\'s, Roderich Moessner, Krzysztof Sacha, and Yizhi You. KG was supported by the Narodowe Centrum Nauki (NCN) grant 2019/32/T/ST2/00413 and the Fundacja na rzecz Nauki Polskiej (FNP) START grant. RL was supported, in part, by the cluster of excellence ct.qmat (EXC 2147, project-id 39085490). PS acknowledges the support of the NCN Sonata Bis grant 2019/34/E/ST3/00405 and the Nederlandse Organisatie voor Wetenschappelijk Onderzoek (NWO) Klein grant via NWA route 2. AK acknowledges the support of \"{O}sterreichische Akademie der Wissenschaften (\"{O}AW) ESQ-Discovery Grant.

\bibliography{fractons}

\end{document}